# Light-induced spiral mass transport in azo-polymer films under vortex-beam illumination

Antonio Ambrosio[1*], Lorenzo Marrucci[1], Fabio Borbone[2], Antonio Roviello[2] & Pasqualino Maddalena[1]

**Affiliation:**
[1]*CNR-SPIN and Dipartimento di Scienze Fisiche, Università degli Studi di Napoli Federico II, Via Cintia, 80126 Napoli, Italy*
[2]*Dipartimento di Scienze Chimiche, Università degli Studi di Napoli Federico II, Via Cintia, 80126 Napoli, Italy*
*Correspondence: antonio.ambrosio@spin.cnr.it

**Abstract:** When an azobenzene-containing polymer film is exposed to a non-uniform illumination, a light-induced mass migration process may be induced, leading to the formation of relief patterns on the polymer free surface. Despite many years of research effort, several aspects of this phenomenon remain poorly understood. Here we report the appearance of spiral-shaped relief patterns on the polymer film under the illumination of focused Laguerre-Gauss beams with helical wavefronts and an optical vortex at their axis. The induced spiral reliefs are sensitive to the vortex topological charge and to the wavefront handedness. These findings are unexpected because the doughnut-shaped intensity profile of Laguerre-Gauss beams contains no information about the wavefront handedness. We propose a model that explains the main features of this phenomenon through the surface-mediated interference of the longitudinal and transverse components of the optical field. These results may find applications in optical nanolithography and optical-field nanoimaging.

**Main Text:** The illumination of a film of a polymer containing azobenzene moieties by means of linearly polarized light, in the UV/visible wavelengths region, leads to the reorientation of the azobenzene units perpendicularly to the light polarization direction[1,2]. This effect is believed to result from the light-induced trans-cis-trans isomerization cycles of the azo-unit, associated with random molecular reorientations, which continue until the rod-like trans-isomer becomes perpendicular to the electric field vector and thus stops absorbing light (although alternative ideas have also been put forward, see, e.g., Refs. 3,4). In 1995, however, another subtler phenomenon has been observed in these materials[5,6], namely a light-induced molecular displacement (or "mass-migration") occurring on the polymer free surface and leading to the formation of stable patterned surface reliefs[7,8,9,10]. A fingerprint of this phenomenon is the conservation of volume: protrusions and hollows appearing on the sample surface preserve the polymer volume on the mesoscopic length scale. These topographical features can be erased by heating the polymer above the glass transition temperature or by illuminating with incoherent uniform light[11]. These writing/erasing phenomena make azo-polymers very attractive for optical data storage applications. The light-induced surface-reliefs could also be used for imaging the electromagnetic field distributions resulting from local illumination by means of near-field sources[12,13,14,15,16] and optical nano-antennas[17]. Recently, the potential advantage of using azo-polymers in the place of sacrificial photoresists in the fabrication of silicon micro- and nano-structures arrays has also been demonstrated[18,19]. A very recent comprehensive review on the science and applications of photoinduced micro and nanostructuring of azobenzene materials is in Ref. 20.

The origin of the mass transport driving force is still debated[20,21,22,23,24,25,26,27,28,29,30]. All currently proposed models however share a common element: the light-induced mass-transport action is linked to the optical field via its intensity-gradients. The relationship is vectorial, as the mass transport appears to occur preferentially in the direction of the electric field[6,7,22,31]. However, apparently nothing in these models predicts a sensitivity of the light-induced mass transport to the wavefront structure of the writing beam. Hence, different optical fields sharing the same polarization and intensity profile would be expected to give rise to the same surface relief patterns. This "naïve" prediction can be tested for example by comparing the patterns induced by two optical vortex beams having opposite wavefront handedness, as we will discuss below.

The concept of "optical vortex" – or "wave-train screw-type dislocation", as it was initially called – was first introduced by Nye and Berry in 1974[32]. The Laguerre-Gauss (LG) beams are the best-known examples of optical modes endowed with an optical vortex located at their beam axis[33]. The optical phase of these modes varies by $2\pi q$ when circling once the beam axis, where $q$ is an integer, positive or negative, called the vortex topological charge. At the axis, i.e. the core of the vortex, the phase is undefined and the optical field must vanish (for nonzero $q$), thus giving rise to the characteristic doughnut shape of the beam intensity cross-section (see Fig. 1). LG beams are characterized by a helical-shaped wavefront, as shown in Fig. 1, and carry so-called orbital angular momentum (OAM)[34,35]. Vortex beams very similar to LG modes, hereafter named LG-like beams, can be conveniently generated from ordinary Gaussian laser beams by diffraction on a suitable pitchfork hologram displayed on a spatial light modulator (SLM)[36,37,38].

Here, we exposed a thin azo-polymer film to focused LG-like vortex beams of varying vortex charge and handedness. The induced spiral reliefs are found to be sensitive to the vortex topological charge and in particular to the wavefront handedness, thus showing that the polymer is responding not only to the intensity distribution of the light field in the focus, but also to its phase. We explain the main qualitative features of this phenomenon with a symmetry-based phenomenological model and discuss the possible underlying microscopic mechanisms at work.

The material used in our experiment is an acrylic polymer bearing the photo-responsive moieties as side chains of the polymeric backbone (see Fig. 2a-inset for the polymer structure). The polymer is spin-coated onto 170 μm thick microscope coverslips. The material absorption spectrum shows a broad maximum in the UV/Visible wavelengths region (Fig. 2a) in accordance with those usually reported for polymers containing azobenzene moieties. At the working wavelength of 532 nm, the polymer film total absorbance is about 5%.

The optical apparatus used for writing the polymer relief structures is shown in Fig. 3. The laser beam, initially in a $TEM_{00}$ Gaussian mode, is injected into an inverted optical microscope and focused on the sample surface by means of an oil-immersion 1.3-numerical-aperture (NA) microscope objective. The laser polarization direction before the focalization is linear, directed as the $y$-axis in the figure. The atomic force microscope (AFM) scanning, reported in Fig. 2b, shows the topographical features obtained on the film surface (polymer-air interface) in this illumination condition. In good agreement with results previously reported in the literature[7,8], we find the formation of two protrusions along the laser polarization direction, with a central hollow. The film thickness is of about 700 nm, the exposure time is of 90 s and the laser power injected into the microscope is 13 μW, corresponding to about 5 μW of estimated exposure power on the polymer. Next, LG-like vortex beams were generated by diffraction on a phase-only SLM and injected in the microscope. Figure 2c reports a photomicrograph of the resulting doughnut intensity profile of the focused beam on the sample (see Methods). This specific beam carries an optical vortex of charge $q = 10$, chosen here because the resulting relief pattern is particularly large and clear. By calibrating the pixel dimension in the micrograph, the inner and outer diameters of the laser spot (including also the outer rings) are measured to be respectively 1.2±0.2 μm and 3.2±0.3 μm at 10% of the maximum intensity. An optical micrograph of the polymer relief pattern induced by this illumination condition is reported in Fig. 2d: a two-arms spiral structure is evident. However, this optical image is affected by light diffraction and the image contrast may result from both the topography modulation and the azo-molecules orientation (local refractive index) in the illuminated region. The actual relief spiral pattern can be observed in the AFM topographical map of the sample surface (Fig. 2e). We stress that no evidence of a spiral structure is present in the light intensity pattern shown in Fig. 2c. Therefore, we are led to conclude that the light-induced mass transport is somehow responding not only to the intensity pattern, but also to the helical structure of the wavefront around the vortex. This conclusion is confirmed when the sign of the optical vortex topological charge is inverted, a change leaving the intensity pattern approximately unmodified (Fig. 2f; we ascribe the small variation in the laser-spot shape to a small residual astigmatism in the optical setup, as high-charge vortex beams are extremely sensitive to astigmatism): the resulting polymer relief pattern is found to have the opposite handedness, as shown in Fig. 2g (optical micrograph) and Fig. 2h (AFM micrograph). Figures 2i and 2j show instead the polarization dependence of the intensity distribution and of the resulting relief pattern, respectively: a 90° rotation of the polarization vector leads to almost no variation of the light intensity distribution, but to a similar 90° rotation of the resulting polymer spiral pattern.

Further details about the observed spiral relief patterns are provided in Fig. 4 with the AFM topographical map of a structure obtained under illumination by a $q = 10$ beam. In particular, the cursors in the image mark the inner (green cursors) and outer (red cursors) diameters of the affected polymer region, which are approximately coincident with the inner and outer diameters of the laser spot shown in Fig. 2c. Hence, it is evident that the material-displacement mainly occurs in the illuminated region whereas the inner part of the structure is made of unperturbed polymer (same height, as the unexposed polymer at the picture's edge). Finally, Fig. 5 shows the dependence of the observed patterns on the light intensity. The most striking observation here is that while the height/depth of the patterns increases with a higher light intensity, the shape of the patterns remains approximately unchanged. Similar results are obtained for a varying exposure time at fixed light intensity.

No model ascribing the mass transport only to the light intensity gradients is capable of describing our results. The spiral arms appearing in the relief pattern must somehow reflect the helical structure of the wavefront, which is lost in the field amplitudes, even separating the three Cartesian field components. One could be tempted to ascribe the spiral relief pattern to a rotational flow induced in the polymer by the absorbed light OAM, but this hypothesis appears to be in contrast with the observation that the pattern keeps the same shape independent of the illumination time or intensity (see Fig. 5), while one would expect instead that the OAM-induced rotational flow makes the spirals wind more for a longer or more intense exposure. Moreover, an order-of-magnitude estimate of the expected induced rotation for our light illumination conditions shows that this effect should be negligible, even assuming a complete light-induced fluidization of the polymer, unless the viscosity dropped to unrealistically small values.

On the other hand, we have found that a double-arm spiral pattern reflecting the wavefront handedness may appear if an interference between the transverse optical field components $E_x$ and $E_y$ and the longitudinal one $E_z$ is involved in the process. Such an interference term is normally absent in isotropic media, but it may become allowed (for a tensorial response of the material) when the rotational symmetry is broken, and in particular in the presence of the medium surface discontinuity, as in our case. Starting from this insight, we have developed a symmetry-based phenomenological theory of the light-induced mass transport process not relying on any specific microscopic model. Our theory is probably oversimplified, as it neglects the viscoelastic couplings in the polymer and the surface tension effects, but nevertheless it is expected to capture the main features of the light driving forces and therefore the qualitative properties of the resulting relief patterns. In the following we present only the main lines of our theory, while more details are given in the Supplementary Information.

The starting assumption of our model is that the optical field induces a mass current **J** that is determined only by the electric field amplitudes and its gradients via a local constitutive relation, to be determined using symmetry constraints. In a two-dimensional (2D) approximation, valid in the limit of a very thin polymer film lying in the *xy* plane, the most general 2D vector that can be built out of quadratic terms in the optical electric field (which are the lowest-order non vanishing ones, after time averaging over the field optical oscillations) and their gradients (which describe the lowest-order non-local dependences in a Taylor expansion) is the following one:

$$\bar{J}_k = C_1 \partial_k \left(E_l^* E_l\right) + C_2 \partial_l \left(E_l^* E_k + E_k^* E_l\right) + C_3 \partial_k |E_z|^2 + \frac{1}{L}\left(C_B E_z^* E_k + C_B^* E_z E_k^*\right) \qquad k,l = x,y \quad (1)$$

where $\mathbf{E} = (E_x, E_y, E_z)$ is the optical electric field in the polymer film (in complex notation), $C_1$, $C_2$, $C_3$ and $C_B$ are constants characterizing the polymer, $L$ is an effective film thickness, $\partial_k$ stands for the partial derivative $\partial/\partial x_k$, and sum over repeated indices is understood. This vector may be taken to represent the polymer 2D mass current in the $xy$ plane (averaged over the thickness $L$ along $z$). A more detailed fully three-dimensional (3D) analysis is provided in the Supplementary Information, where it is also established that $C_1$, $C_2$ and $C_3$ are real-valued quantities linked to bulk processes, while $C_B$ has a dominating surface contribution (so that the resulting $z$-averaged current scales as $1/L$) and in general it might also be complex-valued. Nevertheless, if the field quadratic terms are taken to originate from electric-dipole absorption effects, then also $C_B$ must be real, and we will make this assumption in the following. Further below, we will analyze the physical meaning of these four mass-current terms.

By exposing the polymer to a constant illumination pattern for a certain time $\tau$, the light-induced mass current will give rise to a pattern $\Delta h(x,y)$ of surface "height" variations across the polymer film. Neglecting all possible relaxation effects counteracting the mass migration (e.g., viscoelastic effects, surface tension etc.) and assuming approximate incompressibility of the polymer, $\Delta h(x,y)$ will be given by the following expression (see Supplementary Information for details):

$$\Delta h(x,y) = -\frac{L\tau}{\rho} \partial_k \bar{J}_k = (c_1 + c_2)\left[\partial_x^2 |E_x|^2 + \partial_y^2 |E_y|^2\right] + c_1\left(\partial_x^2 |E_y|^2 + \partial_y^2 |E_x|^2\right)$$
$$+ c_2 \partial_x \partial_y \left(E_y^* E_x + E_x^* E_y\right) + c_3 \left(\partial_x^2 |E_z|^2 + \partial_y^2 |E_z|^2\right) + c_B \left[\partial_x \text{Re}\left(E_z^* E_x\right) + \partial_y \text{Re}\left(E_z^* E_y\right)\right] \quad (2)$$

where $\rho$ is the mass density and we set $c_1 = (L\tau/\rho) C_1$, $c_2 = (2L\tau/\rho) C_2$, $c_3 = (L\tau/\rho) C_3$, and $c_B = (\tau/\rho) C_B$.

Figure 6, from 6a to 6i, reports the simulated distributions, at the focal plane of the objective, of each term appearing in Eq. (2), for a $q = 5$ vortex beam, linearly polarized along the $y$-axis (see Methods for details about the simulations). We choose the $q = 5$ example instead of $q = 10$ here because its smaller overall extension allows using a smaller spatial scale in the images of the predicted relief patterns, thus showing greater detail. Figure 6j shows the total $\Delta h(x,y)$ distribution predicted for the exposed polymer surface as obtained by combining all terms (from Fig. 6a to Fig. 6i), as stated by equation (2). In this figure and in the following ones, we set $c_1 = 0$, $c_2 = 1$, $c_3 = 0$, $c_B = 8$ ($c_2/\lambda$). Indeed, in our system $c_1$ and $c_3$ are found to be negligible with respect to $c_2$ when comparing the theoretical pattern predicted by Eq. (2) with the observations in the case of a linearly polarized Gaussian laser beam with $q = 0$ (see Supplementary Information). Therefore, only the ratio $c_B/c_2$ was adjusted in order to best reproduce the qualitative experimental features of the spiral-shaped structures induced by vortex beams. The resulting pattern $\Delta h(x,y)$ (Fig. 6j) is in very good qualitative agreement with the experimental topographical distribution reported in Fig. 4. In particular, as discussed above, the terms that give rise to the spiral structure are those proportional to the constant $c_B$, corresponding to the interference between the transverse components $E_x$ and $E_y$ and the longitudinal one $E_z$, shown in Figs. 6h and 6i. Furthermore Figs. 6k, 6l, and 6m show the distributions obtained when $q = -5$. In fact, whereas terms from Fig. 6a to Fig. 6g are identical for both $q = 5$ and $q = -5$, terms of Fig. 6h and 6i are modified into those reported in Fig. 6k and 6l respectively. The total relief pattern predicted for $q = -5$ is given in Fig. 6m, which shows the same handedness inversion effect of the spiral arms that we have observed in the experiment (Fig. 2h). Moreover, since in the simulation the only consequence of rotating the polarization direction by 90° is to swap $x$ and $y$-components (plus a sign change), it is clear that our model accounts also for the polarization dependence observed in Fig. 2j.

It is worth noting that the terms displayed in Figs. 6h and 6i (or 6k and 6l), corresponding to the interference between transverse and longitudinal field components, vanish identically (or are negligible) in the case of light beams that have not a vortex structure or a helical wavefront. To our knowledge, these include all the illumination patterns used in the previously reported experiments. This is the reason why the wavefront-sensitive mass transport effects reported here have not been noticed before.

We have already shown that our model can qualitatively explain the two-arms spiral shape of the polymer relief pattern, its handedness dependence on the optical wavefront handedness, and its polarization dependence. In addition, Figure 7 reports the evolution of the twisting of the experimental and theoretical polymer structures as a function of the vortex beam topological charge $q$. The qualitative agreement between experiment and theory is evident. More in detail, by increasing $q$ from 1 to 10, the experimental and theoretical patterns both show an increasing transverse extension, a decreasing height and depth of the reliefs, and an increasing rotation angle of the spiral arms with respect to the $y$-axis. However the model does not fully account for the somewhat different behavior of the topographic maxima and minima, which are asymmetric in the experiments and more symmetric in the simulations. This and other more quantitative discrepancies are presumably due to the occurrence of viscoelastic and surface-tension stresses in the polymer during the experiment, opposing the material-displacement, which are not accounted for in our simplified theory. A more quantitative model will be developed in future work, by combining the photoinduced current given in Eq. (1) with viscous forces so as to write a generalized Navier-Stokes equation[25], to be complemented with the appropriate boundary conditions for the polymer flow and surface tension effects.

Above, we introduced a symmetry-based phenomenological theory for the light-induced mass transport that needs no assumptions about the underlying microscopic mechanism. From a practical point of view, this is an advantage of our approach, as there is no consensus yet about the correct microscopic picture[20]. Our phenomenological theory is powerful enough to make specific predictions about the induced patterns, particularly after determining the value of its four phenomenological coefficients $C_1$, $C_2$, $C_3$ and $C_B$, which in our lowest-order approximation may only depend on the material properties. The predicted 2D current given in Eq. (1) presents four corresponding separate terms, whose phenomenological physical meaning can be given in terms of the resulting mass-transport effects. For example, the current term in $C_1$ corresponds to a mass migration along the gradient of the total transverse intensity ($\propto |E_x|^2 + |E_y|^2$), hence it is just driving the polymer molecules out of the bright regions (assuming $C_1>0$), irrespective of the polarization direction. In our theory, this term is the only possible cause of the relief gratings obtained for s-s polarized two-beam interference (see Supplementary Information), which are usually found to be much less pronounced than the gratings obtained for other polarization combinations[20] (this shows that $C_1$ is small in most materials). The term in $C_3$ gives rise to the same intensity-gradient effect, but accounting for the additional intensity associated with the longitudinal field $\propto |E_z|^2$ (and for an isotropic 3D polymer response we must have $C_1=C_3$, as shown in the Supplementary Information; so also $C_3$ is likely small). The term in $C_2$ is instead causing the polarization-sensitive anisotropic mass transport, by inducing motion only along the direction of the electric field. This term is for example what makes the relief gratings induced by p-p polarized two-wave interference much more pronounced than s-s ones (see, e.g., Ref. 20), what causes the directional fluidization reported in Ref. 31, and what gives rise to the two lobes appearing after illumination by a single linearly-polarized Gaussian beam (see, e.g., Fig. 2a and Ref. 7). Thus, former investigations concur in indicating that the term in $C_2$ is the dominating one, consistent with the assumptions $C_1 = C_3 = 0$ used in our simulations. Finally, the newly-predicted current term in $C_B$ is what induces the spiral transport effect, being sensitive to the wavefront handedness via the interference of the longitudinal and transverse field components.

Let us now briefly discuss the possible physical meaning of these four current terms in connection with one of the microscopic mechanisms that have been proposed in the literature to explain the mass migration. Such microscopic models are reviewed in Ref. 20 (see, e.g., Table 1), with the conclusion that no single model is presently capable of explaining all observed features of the photoinduced phenomena in azobenzene materials. Probably a realistic model must combine several effects, as for example attempted in the numerical simulations reported in Ref. 30. Nevertheless, we have developed an analytical microscopic model based on the light-induced anisotropic diffusion (or random walk) of the molecules as the main underlying mechanism to explain the mass migration[24,28,29], which is also one of the key ingredients of the simulations reported in Ref. 30. We will publish the details of this model elsewhere, but we can anticipate here that its results are entirely consistent with those of our phenomenological theory. In the framework of this specific microscopic model, the mass-current terms with coefficients $C_1$, $C_2$ and $C_3$ appearing in our Eq. (1) are associated with the light-driven anisotropic molecular diffusion of azobenzene moieties occurring in the polymer bulk, with the diffusion along the polarization direction being strongly favored by the more likely excitation of the azomolecule chromophores aligned along the electric field direction. The $C_B$ term is associated with a similar light-driven anisotropic diffusion, but combined with an enhanced mobility of the azo molecules lying close to the polymer boundaries (in particular at the polymer surface) and with the obvious additional constraint that the molecules cannot leave the polymer medium. We stress, however, that this specific microscopic interpretation of the light-induced mass currents must be regarded as tentative, at this stage.

In conclusion, we have shown that a solid film made of an azobenzene-containing polymer is sensitive to the helical wavefront handedness of a doughnut laser beam, so as to develop spiral-shaped relief patterns responding to the wavefront handedness and topological charge. We ascribe this phenomenon to the action of an unusual, perhaps unprecedented, interference between longitudinal and transverse optical field components, made possible by the symmetry breaking taking place at the polymer surface. Our findings open new possibilities in azo-polymer-based micro- and nano-lithography[18,19], allowing to design more complex patterns by exploiting the light wavefront as an additional control handle. Furthermore, the insights provided by our model, by advancing our understanding of the link between the driving optical field and the resulting polymer topographical patterns, will contribute to exploiting the light-induced mass-migration phenomenon for the non-optical nano-imaging of near-field electromagnetic sources and scattering elements[20].

**Materials & Methods**

The reagents for the synthesis were purchased from Aldrich and used without further purification. $^1$H nuclear magnetic resonance (NMR) spectra were recorded on a Varian XL 200 MHz and chemical shifts are reported as δ values (ppm) relative to internal Me$_4$Si. Differential scanning calorimetry (DSC) measurements were performed on an indium-calibrated Perkin–Elmer Pyris 1 apparatus, under a dry nitrogen atmosphere with a temperature scanning rate of 10 °C/min. UV/Vis absorption spectra were recorded with a Jasco V560 spectrophotometer in chloroform.

The photoresponsive monomer is an azobenzene with symmetric distribution of alkoxy substituents and a terminal acrylic group, (E)-2-(4-((4-methoxyphenyl)diazenyl)phenoxy)ethyl acrylate. The monomer was synthesized according to a method reported in the literature[39,40]. The polymer was obtained by radical polymerization of the monomer in solution according to the following procedure. (E)-2-(4-((4-methoxyphenyl)diazenyl)phenoxy)ethyl acrylate (1.00 g, 3.06 mmol) and 2,2′-azobis(2-methylpropionitrile) (0.0100 g, 6.09·10$^{-2}$ mmol) were dissolved in 4 mL of N,N-dimethylformamide in a vial that was sealed under vacuum after three "freeze and thaw" cycles. The solution was kept at 70°C for 48 hours and then poured into 100 mL of methanol. The polymer was filtered, washed with methanol, dissolved in chloroform and precipitated in hexane. $^1$H NMR (CDCl$_3$, 200MHz): δ 1.20 (s); 1.68 (s); 1.97 (s); 2.48 (s); 3.82 (s); 4.02 (s); 4.29 (s); 6.89 (s); 7.78 (s). Phase sequence: G 67°C N 113°C I (G: glass, N: nematic, I: isotropic). UV-Vis: $\lambda_{max}$ 356 nm.

The $^1$H NMR spectrum of the polymer shows broad singlets, no acrylate resonances and four signals in the aliphatic region, due to a significant amount of insertion errors and regio-irregularity generated in the experimental conditions. Thin films were deposited from 1,1,2,2-tetrachloroethane solution by filtering on 0.2 μm teflon filters and spin coating onto glass substrates. By means of this technique the polymer can be obtained in the amorphous state, which is stable at room temperature and the nematic phase only arises upon annealing. This behavior allowed the realization of photoisomerization experiments on unstructured samples. The endothermal transition at 113°C corresponds to isotropization of the nematic phase. DSC scans were also performed after isothermal treatments at 90°C for 0.5, 1 and 2 hours. The curves show the same peak area at 113°C and hence unchanged heat of isotropization, thus indicating the nematic phase to form rapidly and completely after annealing the amorphous phase or cooling down from the isotropic region.

The holographically-produced vortex beams are obtained by means of a phase-only Spatial Light Modulator (Pluto, by HOLOEYE Photonics AG). The input Gaussian beam is first spatially expanded in order to fill the SLM active area (1080×1920 pixels, 8μm wide each). The beam diffracted (in reflection) by the SLM is then propagated towards the microscope by means of two consecutive lens-systems in telescopic configuration (see Fig. 3). The focal lengths of the lenses are chosen so as to achieve the correct magnification of the beam size before injecting it into the microscope objective. This is an inverted microscope where the image is formed by means of the same microscope objective used to focus the laser beam at the sample-air interface. The polymer film is spin-coated on glass coverslips and matching oil is usually added in between the microscope objective and the coverslip free surface. The laser-spot photomicrographs shown in Fig. 2 are obtained by mounting a digital camera in the place of the eyepieces. In order to avoid the typical aberrations in imaging by reflection from a dielectric interface when using high-numerical-aperture objectives (see, e.g., Ref. 41), these photos were taken with a metallic mirror inserted on the top of the coverslip in the place of the polymer film sample (in this case, matching oil is also added in between the coverslip and the mirror surface). The multiple concentric rings observed for the intensity profile of a vortex beam (Fig. 2) are to be ascribed mainly to diffraction arising in the microscope objective, since we are in an overfilling geometry.

The AFM images have been obtained using an Atomic Force Microscope (XE-100 by Park Systems Corp.) working in non-contact mode. The image analysis has been performed using standard software provided by the same company.

The focalization of a linearly polarized beam by means of a high numerical aperture microscope objective results in non-negligible field components along the other two Cartesian directions[42]. The complex quantities $E_x$, $E_y$ and $E_z$ of the electromagnetic field at the focal plane of the objective are obtained by using the angular spectrum representation of the field refracted by the lens combined with the stationary phase method for the evaluation of non-evanescent field components. The resultant quantities are then calculated by using numerical integration on the collection solid angle, for each of the points in the simulated area, in accordance to the work of Richards and Wolf[43,44] for aplanatic lenses and the treatment reported, for instance, in reference 41. An example of the results of these calculations is reported in Fig. S1. The derivatives appearing in equation (2) are also numerically evaluated by approximating the spatial derivatives of the field components with the incremental ratios, with a spatial step of 0.01 times the light wavelength. This guarantees a good approximation of both the continuous fields and derivatives. An example of the results obtained in the case of a LG-like beam is shown in Fig. 6 and, for an ordinary Gaussian beam, in Fig. S2.

**Acknowledgemets**
The research leading to these results has received funding from the European Community, 7[th] Framework Programme, under grant n. 264098 – MAMA and under grant n. 255914 – PHORBITECH, the latter within the Future Emerging Technologies FET-Open programme.

**Author contributions**
A.A. conceived and conducted the experiments and performed the numerical simulations. L.M.

developed the theoretical modeling. F.B. and A.R. synthesized the polymer and provided the samples. A.A., L.M. and P.M. discussed the experimental data and the simulation results and wrote the paper, with some input from the other authors. P.M. initiated and directed the project.

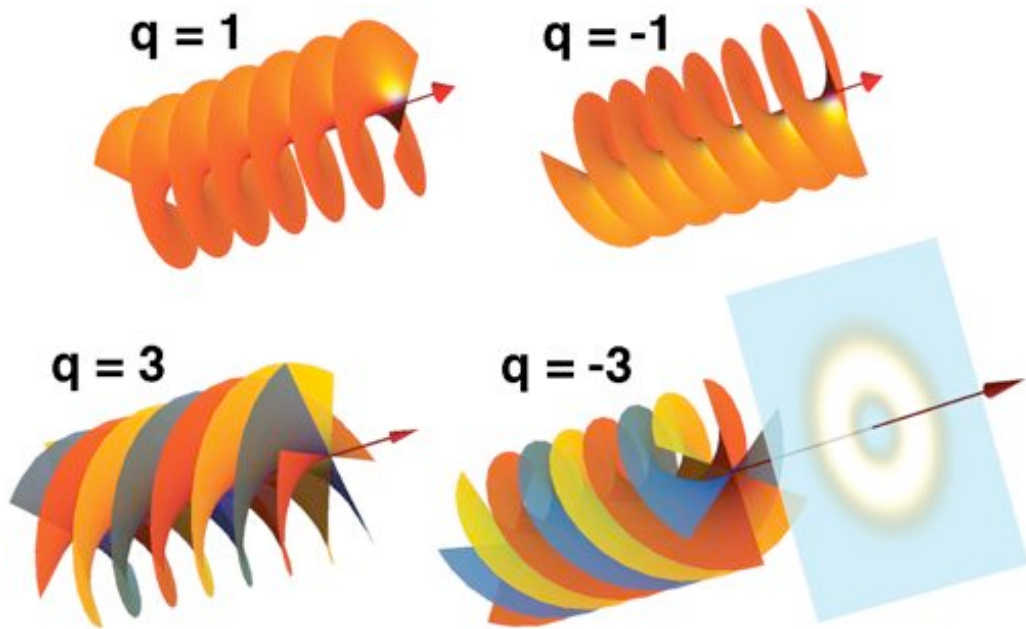

**Figure 1.** Schematics of the vortex beam optical field employed in the present work. Shown are the wavefront helical structures for vortex topological charges $q = \pm 1$ and $q = \pm 3$ (in the latter case, the wavefront is composed of three intertwined helical surfaces, here shown in different colors for clarity) and, in the last example, the associated doughnut-shaped transverse intensity distribution.

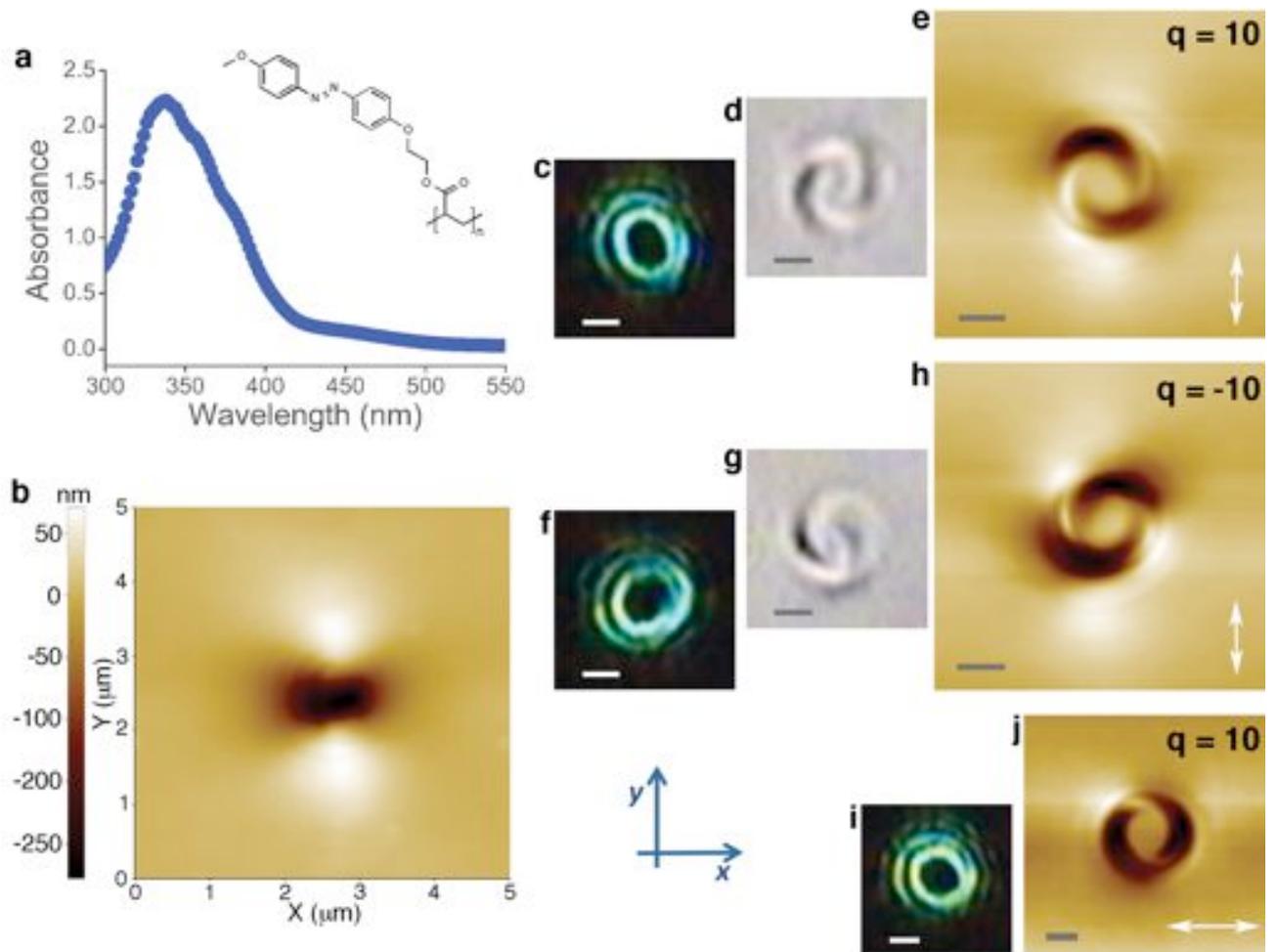

**Figure 2. a**, Absorption spectrum of the azo-polymer used in this work. Inset: polymer structure. **b**, Surface-relief pattern induced by a focused Gaussian beam (topological charge $q = 0$). The material displacement leads to a two-lobed pattern oriented along the beam polarization direction ($y$-axis), as already reported in previous works[7,8]. **c**, Optical micrograph of the intensity pattern of a LG-like vortex beam having topological charge $q = 10$, focused by a 1.3 NA oil-immersion microscope objective on the surface of a coverslip where a metallic mirror is inserted in the place of the sample. **d**, Optical micrograph of the polymer surface-relief pattern generated by the focused $q = 10$ vortex beam (linearly polarized along the $y$-direction). The image is taken by means of the same objective used to illuminate the sample. **e**, AFM image of the same surface-relief pattern as in **d**. **f**, Optical micrograph of the intensity pattern when the vortex handedness is inverted (topological charge $q = -10$; still linearly polarized along the $y$-direction). **g**, Optical micrograph of the polymer surface-relief pattern induced after inverting the vortex handedness. **h**, AFM image of the same structure as in **g**. **i**, Optical micrograph of the intensity pattern obtained for a vortex beam with $q = 10$, when the light polarization direction is rotated by 90° (so as to be parallel to the $x$-axis). **j**, AFM surface-relief pattern corresponding to the illumination conditions of **i**: the two arms of the spiral are rotated by 90°, as compared to the pattern shown in **e**. The white arrow in panels **e**, **h** and **j** indicates the polarization direction of the light. The scale-bars in panels **c-j** correspond to 1 μm.

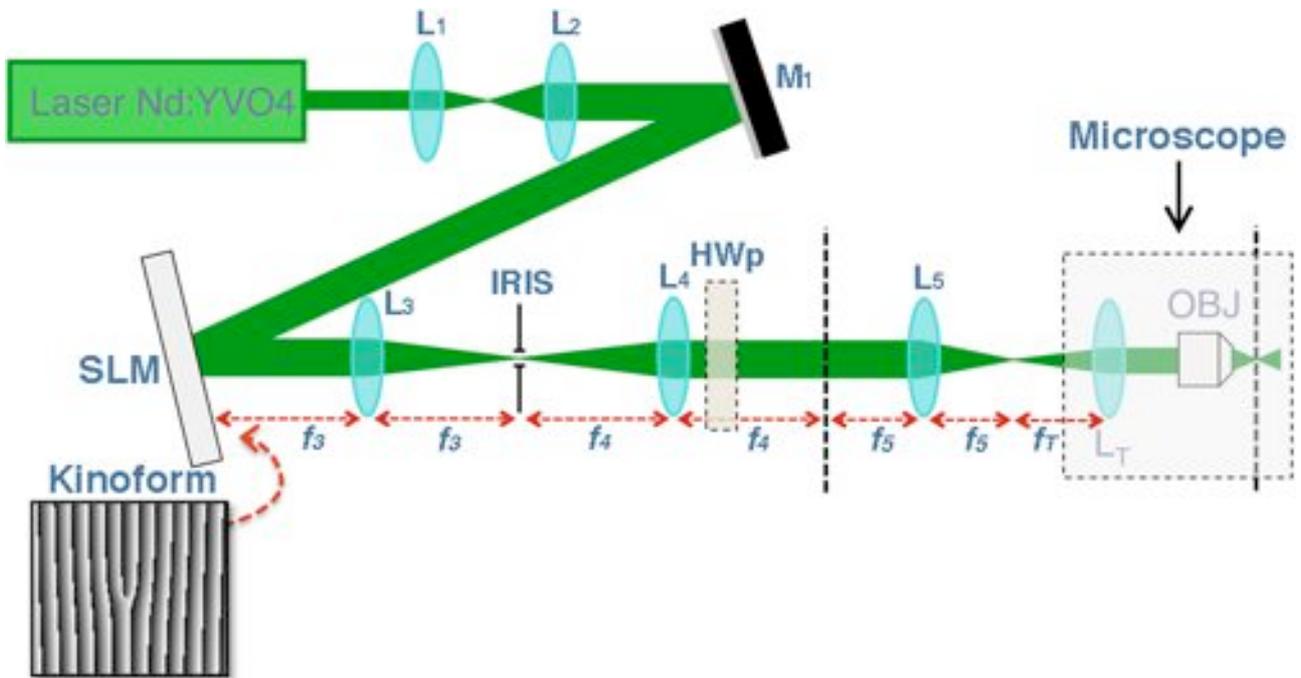

**Figure 3.** The laser beam (from a Nd:YVO$_4$ continuous-wave frequency-duplicated laser), after a beam-expander (lenses L$_1$ and L$_2$), is sent onto a computer-controlled SLM which is programmed for visualizing a pitchfork hologram (kinoform) generating the desired LG-like beam with a prescribed topological charge. Next, the first-order diffracted beam is selected via an iris located in the focal plane of a lens (L$_3$). After recollimation (lens L$_4$), the beam is sent through a half-wave plate (HWp) for rotating the input polarization and finally imaged by external lens L$_5$ and the internal lens system of the microscope (including tube lens L$_T$ and the microscope objective OBJ) to the sample plane, positioned at the polymer surface. The dashed lines mark the image planes reproducing the optical field after SLM diffraction.

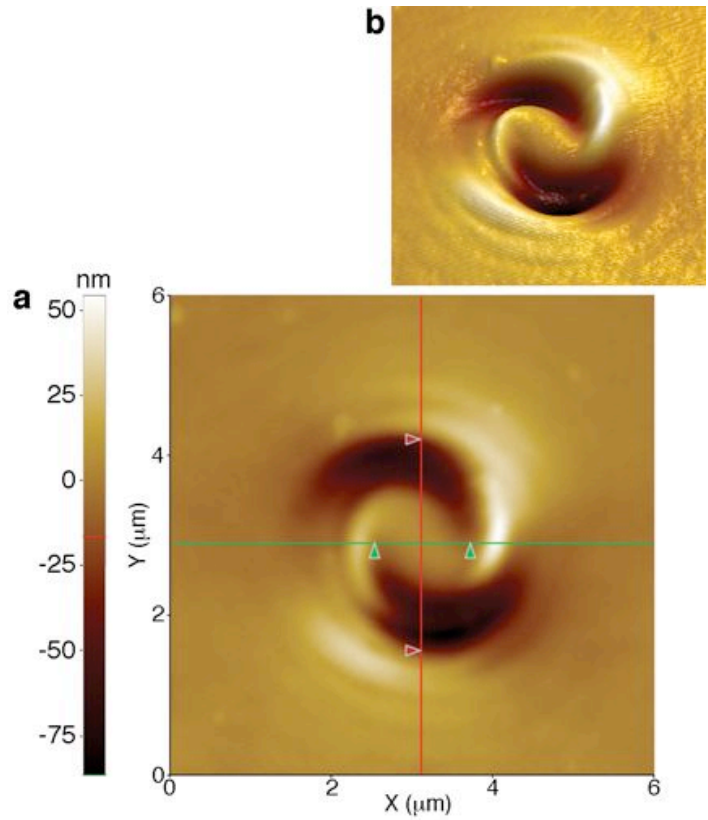

**Figure 4. a**, Two-dimensional AFM image of the topographical structure obtained at the sample surface when the polymer is illuminated by a focused $q = 10$ vortex beam. The distances between the two green and the two red cursors, taken along the corresponding lines passing through the structure center, are 1.2 µm and 2.7 µm, respectively. These values are consistent with the inner and outer diameters of the laser spot shown in Fig. 2 c. **b**, Three-dimensional rendering of the same AFM map.

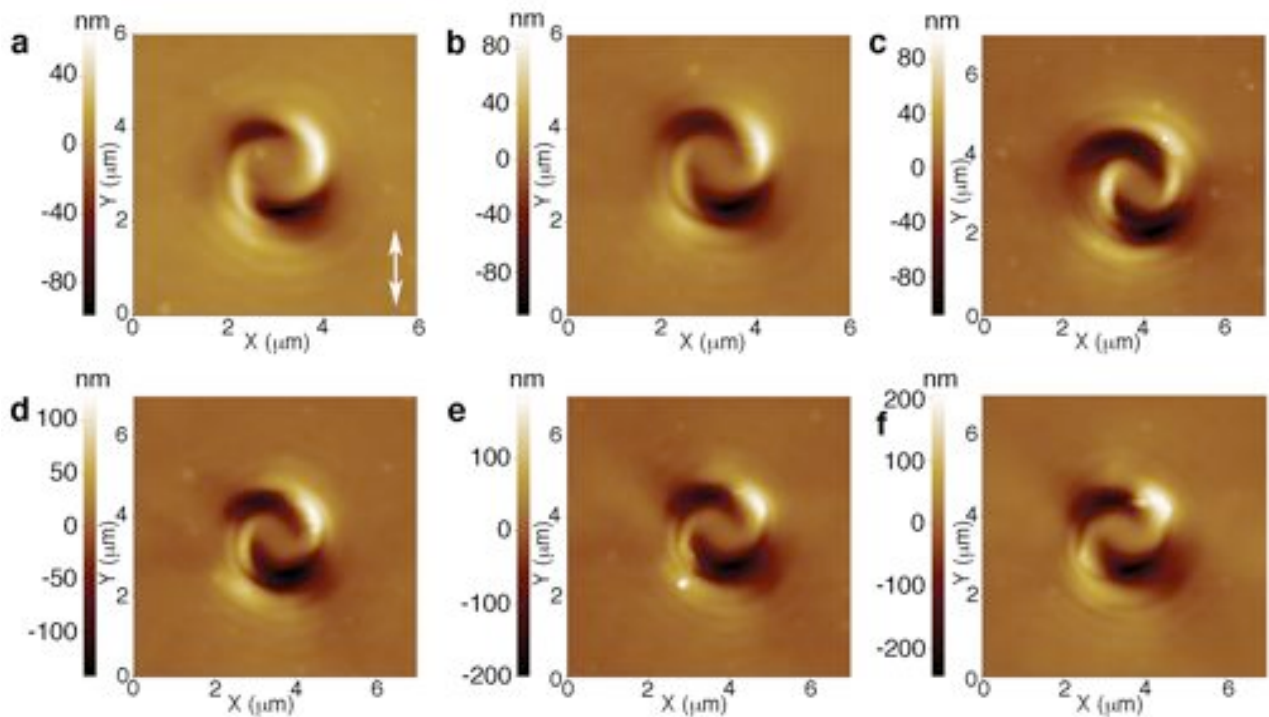

**Figure 5.** AFM images of the topographical structures obtained for varying illumination intensity and fixed time of exposure (and polarization direction), for topological charge $q = 10$. The white arrow indicates the polarization direction. Different panels correspond to different values of the laser power injected in the microscope: **a**, 15 µW; **b**, 18 µW; **c**, 21 µW; **d**, 29 µW; **e**, 41 µW; **f**, 54 µW. Similar results are obtained for varying time of exposure at fixed intensity.

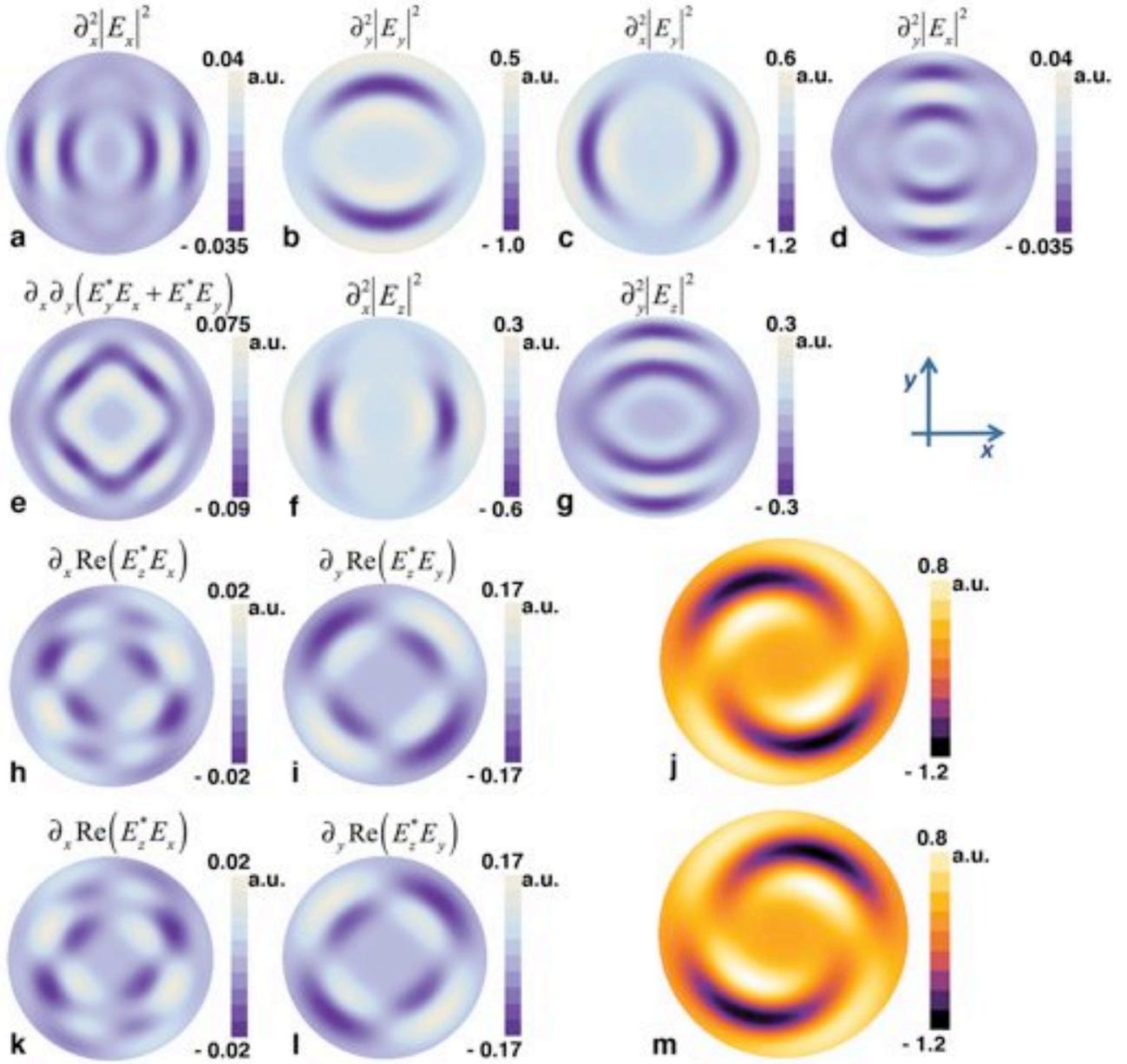

**Figure 6. a**, Contribution to the surface-relief pattern resulting from the second-derivative term $\partial_x^2|E_x|^2$ appearing in our model, based on the simulations of the optical field at the focus, for a tightly-focused $q = 5$ vortex beam, linearly polarized along the $y$-axis. The relief scale is in arbitrary units. The simulated area is a circle of four times the light wavelength in diameter. **b**, **c**, **d**, **e**, **f**, **g**, **h**, **i**, Same as in **a**, for the other field-derivative terms appearing in our model and displayed above each panel. **j**, Total surface-relief pattern obtained by combining all terms as in Eq. (2), with the constants' values set to: $c_1 = 0$; $c_2 = 1$; $c_3 = 0$; $c_B = 8$ $(c_2/\lambda)$. **k**, **l**, **m**, Same as for **h**, **i** and **j**, respectively, when the vortex beam charge is inverted to $q = -5$ (all other terms remain unchanged).

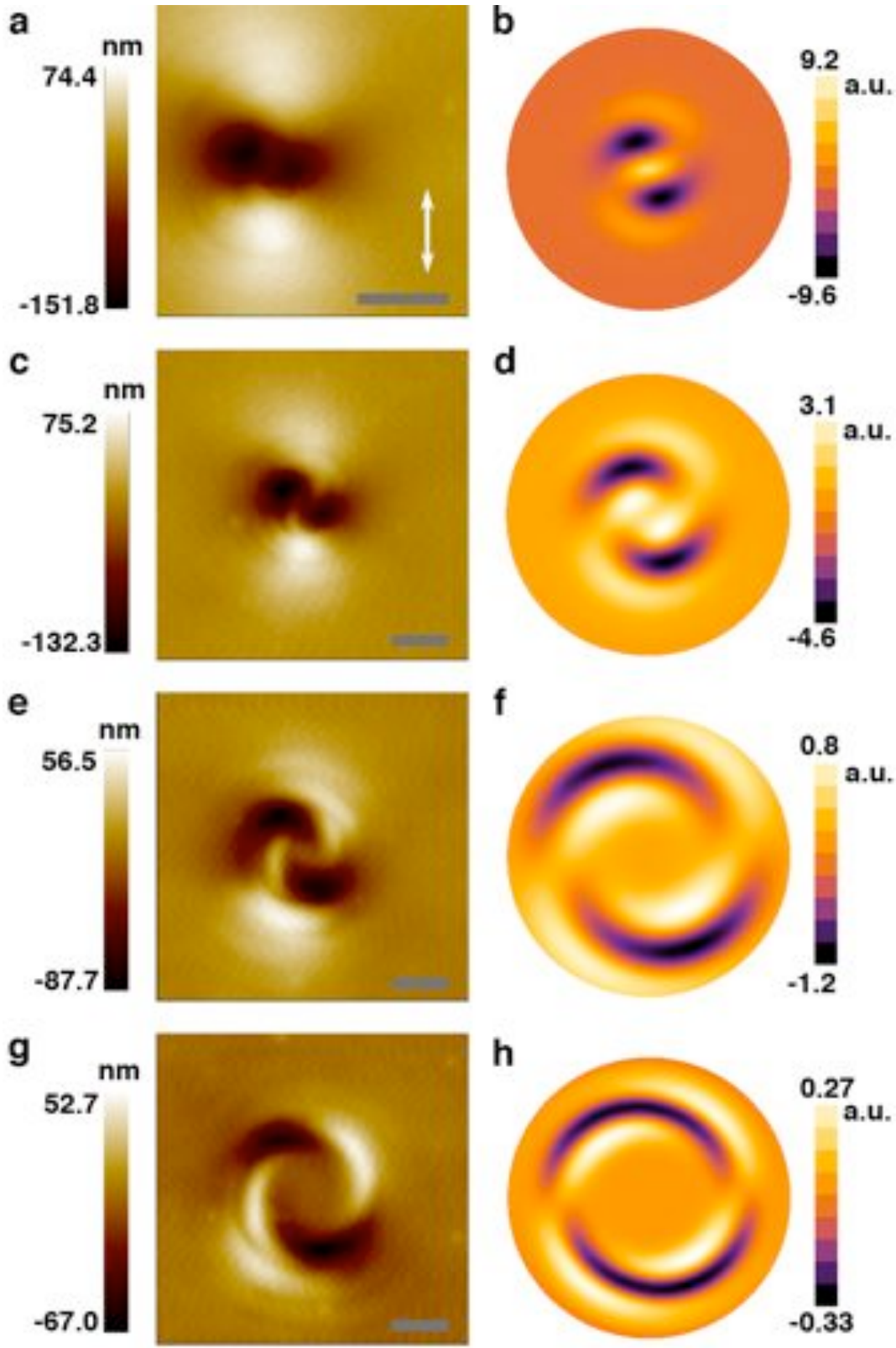

**Figure 7.** The topological charge increases when moving from the top to the bottom of the figure. **a**, Topographical AFM map of the pattern obtained with a $q = 1$ vortex beam, linearly polarized as indicated by the white arrow. **b**, Corresponding simulation based on our model. **c** and **d**, same as for panels **a** and **b**, respectively, but in the case $q = 2$. **e** and **f**, case $q = 5$. **g** and **h**, case $q = 10$. The scale-bars of all the AFM maps correspond to 900 nm. The simulated area in **b**, **d** and **f** has a diameter of 4 times the wavelength, whereas in **h** it has a diameter of 7.3 times the wavelength.

Supplementary Information for

# Light-induced spiral mass transport in azo-polymer films under vortex-beam illumination

Antonio Ambrosio, Lorenzo Marrucci, Fabio Borbone, Antonio Roviello, Pasqualino Maddalena

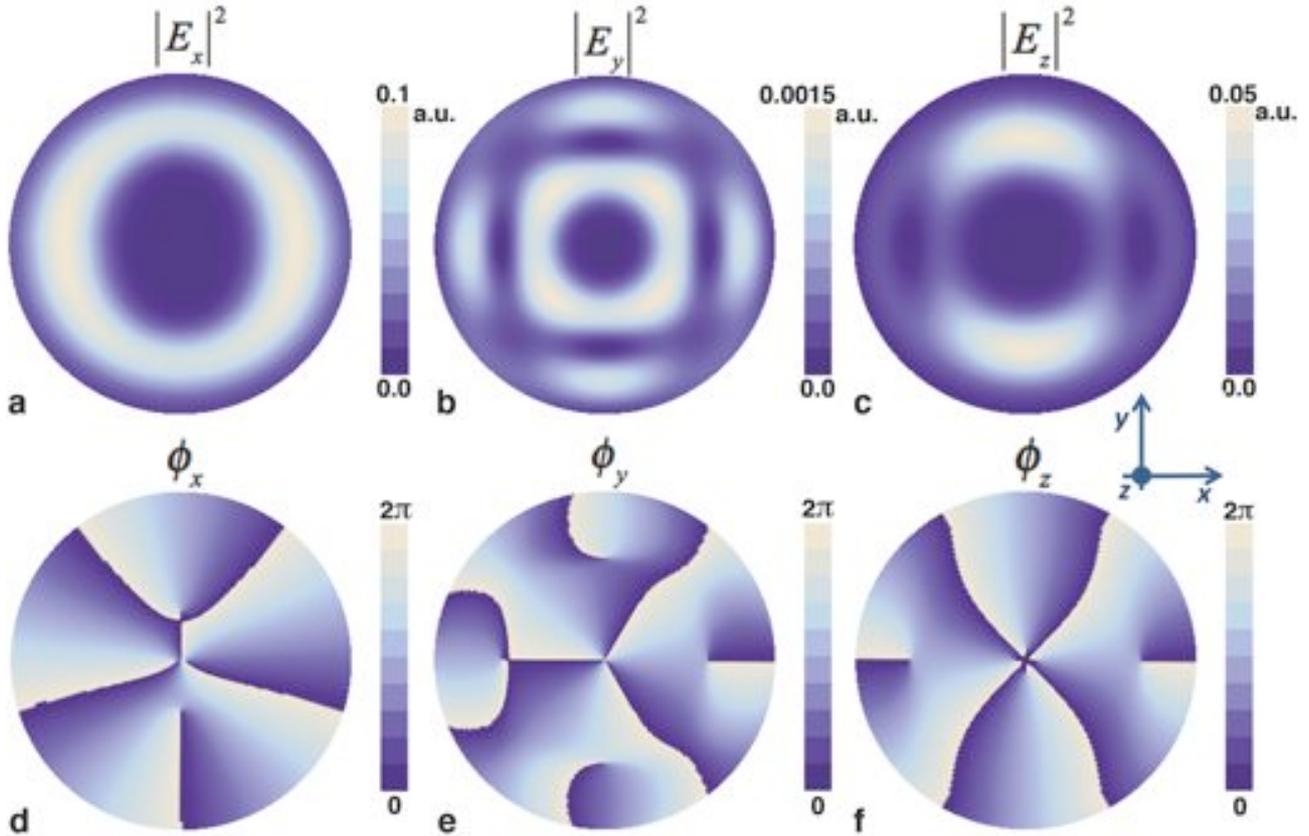

**Figure S1. a**, Simulated two-dimensional distribution of the $|E_x|^2$ component of a $q = 5$ Laguerre-Gauss beam at the focal plane of a oil-immersion 1.4 NA microscope objective (values in arbitrary units). The beam is linearly polarized along the *x*-azis. **b**, **c**, The same as **a** for the components $|E_y|^2$ and $|E_z|^2$ of the electromagnetic field. **d**, Optical phase around the beam propagation axis (*z*) for the $E_x$ component of a $q = 5$ Laguerre-Gauss beam at the focal plane. **e**, **f**, The same as for **d** for the $E_y$ and $E_z$ components, respectively. The simulated area has a diameter of four optical wavelengths.

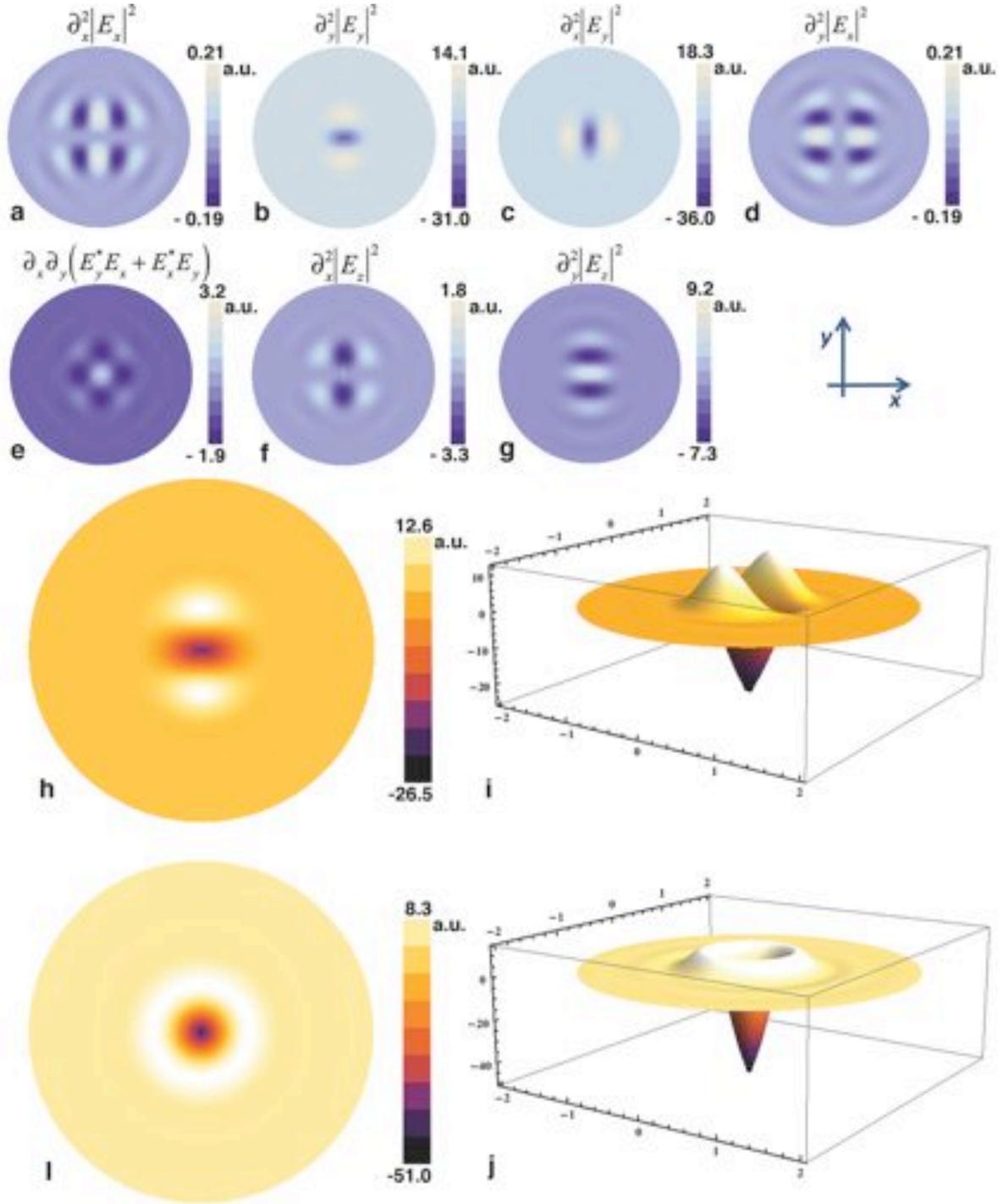

**Figure S2. a**, Simulation of the two-dimensional distribution at the focal plane for the quantity $\partial_x^2|E_x|^2$ in the case of a tightly-focused Gaussian laser beam, linearly polarized along the *y*-axis. **b**, **c**, **d**, **e**, **f**, **g**, The same as in **a** for the quantities reported above each distribution. **h**, Two-dimensional map of the height variation, $\Delta h(x,y)$, of the topographical structure obtained from our model (Eq. 2 of the main Article) by setting the constants' values to: $c_1 = 0$; $c_2 = 1$; $c_3 = 0$. **i**, Three-dimensional representation of the same distribution reported in **h**. The material displacement occurs mainly along the polarization direction. **l**, The same as for **h** in the case of illumination by means of a circularly polarized Gaussian beam. **j**, Three-dimensional representation of the same distribution reported in **l**. The simulated area has a diameter of four wavelengths.

**Phenomenological model of light-induced mass transport.** Let us consider a polymer thin film deposited on a rigid substrate and initially extending in the region comprised between the plane $z = 0$ (polymer-substrate interface) and the plane $z = L$ (polymer free surface). Actually, it is possible that the polymer region in which the light-induced mass migration occurs does not correspond to the entire polymer thickness, in which case $L$ will represent the effective thickness of this "mobile" region of the polymer and the plane $z = 0$ will correspond to an inner polymer layer at which no light-induced motion can take place. After exposure to light, the polymer develops surface reliefs, which can be described by the function $h(x,y)$ giving the new $z$-coordinate of the free surface, or equivalently by the height variations $\Delta h(x,y) = h(x,y) - L$. We assume that these surface reliefs arise as a consequence of light-induced mass transport, as described by a mass-current-density vector **J** determined by the optical field. We neglect possible additional contributions to the mass-current-density, such as visco-elastic interactions between adjacent moving regions of the polymer. This mass transport acts on the polymer by varying the local mass density $\rho$, which in turn determines a deformation of the polymer film via its elastic response. However, here we assume the validity of a simplified limit in which the light-induced mass density variations are exactly balanced by a local expansion (or contraction) of the polymer, so as to return to the initial equilibrium density, according to the following law (incompressibility approximation):

$$\frac{\Delta V}{V} = \partial_i u_i = \frac{\Delta \rho_{\text{light}}}{\rho} \quad (S1)$$

where $u_i$ is the elastic displacement vector and sum over repeated indices is assumed. Moreover, we assume that the film thickness $L$ is much smaller than the imposed lateral variations of $\Delta\rho_{\text{light}}$ along $x$ and $y$ and hence the derivatives of $u_i$ with respect to $x$ and $y$ can be neglected in Eq. (S1) (this assumption will be valid if the effective thickness $L$ is much smaller than the laser spot-size). These assumptions yield

$$\partial_z u_z = \frac{\Delta \rho_{\text{light}}}{\rho} \quad \Rightarrow \quad u_z(L) = h(L) - L = \Delta h = \int_0^L \frac{\Delta \rho_{\text{light}}}{\rho} dz \simeq \frac{1}{\rho}\int_0^L \Delta \rho_{\text{light}} \, dz = \frac{L}{\rho}\overline{\Delta \rho} \quad (S2)$$

where $\overline{\Delta \rho}$ is the light-induced density variation averaged along $z$ over the entire effective thickness and we have used the fact that the substrate is rigid, so that $u_z(0) = 0$. The link between the light-induced mass density variations and the light-induced current density **J** is provided by the mass continuity equation:

$$\frac{\partial \rho_{\text{light}}}{\partial t} = -\partial_i J_i \quad \Rightarrow \quad \frac{\partial \overline{\rho}}{\partial t} = -\frac{1}{L}\int_0^L \partial_i J_i \, dz = -\partial_k \overline{J}_k - \frac{1}{L}[J_z(L) - J_z(0)] = -\partial_k \overline{J}_k \quad (k = x, y) \quad (S3)$$

where we have introduced the averaged lateral currents $\overline{J}_k$ with $k$ spanning only the two transverse coordinates $x$ and $y$, while the current $J_z$ is assumed to vanish at the bounding surfaces because there can be no mass transport out of the polymer. Combining Eqs. (S2) and (S3) and assuming an exposure time $\tau$ during which the light-induced current density is taken to remain constant, we obtain the following final expression of the reliefs which will be used in the following:

$$\Delta h(x,y) = -\frac{L\tau}{\rho}\partial_k \overline{J}_k \quad (k = x, y) \quad (S4)$$

We can see from this equation that, in order to determine the surface reliefs, within the present approximation we only need to know the lateral current density induced by light, averaged over the polymer thickness. Precisely, we need to write a constitutive equation that gives the current density **J** resulting from a given applied optical field, as described by the electric field **E** and magnetic field **B**. We can exclude a linear dependence of **J** on **E** or **B**, as for a quasi-monochromatic field this term would average to zero. Therefore, the lowest-order dependence must be quadratic in the field components. Moreover, we can reasonably assume that the material response (being related with light absorption) is sensitive to the electric field **E** only, and not to the magnetic field, as the contribution of the latter to absorption is typically very weak. In particular, in the polymer bulk we can write the following fully general 3D constitutive equation, which assumes only isotropy of the polymer (at equilibrium), a lowest-order quadratic response on the optical field, and a lowest-order linear dependence of the field gradients:

$$J_i = C_1 \partial_i \left( E_j^* E_j \right) + C_2 \partial_j \left( E_j^* E_i \right) + C_2^* \partial_j \left( E_i^* E_j \right) \quad i = x, y, z \quad (S5)$$

where $C_1$ and $C_2$ are two constants characteristic of the material and Maxwell's equation $\partial_i E_i = 0$ was used to remove possible additional terms. Notice that $C_1$ must be real while $C_2$ in general could be complex. However, the effect of the current is determined only by its divergence $\partial_i J_i$, in which only the real part of $C_2$ survives, as can be seen by a direct calculation. Therefore, we can take also $C_2$ to be real without loss of generality. At the polymer surfaces we may have an additional surface-enhanced contribution to the lateral (2D) current of zero-order in the field gradients (hence possibly stronger):

$$J_k = \left( C_s E_z^* E_k + C_s^* E_z E_k^* \right) \delta(z - L) + \left( C_i E_z^* E_k + C_i^* E_z E_k^* \right) \delta(z) \quad k = x, y \quad (S6)$$

where $C_s$ ($C_i$) is a (generally complex) constant characteristic of the polymer surface (interface with substrate) and we have introduced Dirac's delta function $\delta(z)$ to represent the surface localization of this extra current. There can be no z component of the surface current because it would imply a flow of mass out of the polymer. In addition, if Eq. (S5) predicts a nonzero z-component of the current at the polymer boundary, we must assume the presence of additional surface-specific effects (related with the polymer cohesion energy) that will balance them. These terms will not contribute to the lateral currents, and therefore we need not find their explicit expression (they act as a constraint). Combining Eqs. (S5) and (S6), and distinguishing explicitly the xy-components and the z-one, we obtain

$$\begin{aligned} J_k &= C_1 \partial_k \left( E_l^* E_l \right) + C_1 \partial_k |E_z|^2 + C_2 \partial_l \left( E_l^* E_k + E_k^* E_l \right) + C_2 \partial_z \left( E_z^* E_k + E_k^* E_z \right) \\ &\quad + \left( C_s E_z^* E_k + C_s^* E_z E_k^* \right) \delta(z - L) + \left( C_i E_z^* E_k + C_i^* E_z E_k^* \right) \delta(z) \quad k, l = x, y \end{aligned} \quad (S7)$$

We now average along z, across the entire polymer thickness L:

$$\begin{aligned} \overline{J}_k &= C_1 \overline{\partial_k \left( E_l^* E_l \right)} + C_1 \overline{\partial_k |E_z|^2} + C_2 \overline{\partial_l \left( E_l^* E_k + E_k^* E_l \right)} + \frac{C_2}{L} \left( E_z^* E_k + E_k^* E_z \right) \Big|_0^L \\ &\quad + \frac{1}{L} \left( C_s E_z^* E_k + C_s^* E_z E_k^* \right)_{z=L} + \frac{1}{L} \left( C_i E_z^* E_k + C_i^* E_z E_k^* \right)_{z=0} \quad k, l = x, y \end{aligned} \quad (S8)$$

It should be noticed that the fourth term in this expression has exactly the same dependence on the fields and the thickness as the fifth and sixth ones, and may be therefore reabsorbed within them by simply redefining the constants $C_s$ and $C_i$. Therefore, we drop the fourth term in the following. Moreover, we make the further assumption that the transverse fields $E_x$, $E_y$ vary only little across the polymer (this corresponds to assuming that $L \ll z_0 = \pi w_0^2 / \lambda$, where $w_0$ is the beam waist radius and $\lambda$ is the wavelength) and therefore replace the averaged fields with their punctual value (at any z within the polymer). Within the same assumption, the two surface and interface terms will give the same effect, and we may therefore collect them by introducing a single new "boundary-related" constant

$$C_B = C_s + C_i \quad (S9)$$

After this last assumption, we obtain our final "phenomenological" expression for the averaged lateral current (corresponding to Eq. (1) of the main manuscript):

$$\overline{J}_k = C_1 \partial_k \left( E_l^* E_l \right) + C_2 \partial_l \left( E_l^* E_k + E_k^* E_l \right) + C_3 \partial_k |E_z|^2 + \frac{1}{L} \left( C_B E_z^* E_k + C_B^* E_z E_k^* \right) \quad k, l = x, y \quad (S10)$$

In the last expression, we distinguished the third term by introducing a separate constant $C_3$, for increasing the generality of our treatment. From Eq. (S8) we have $C_3 = C_1$ due to 3D isotropy of the bulk polymer, but the two constants might actually become slightly different if there is some anisotropy effect along z. Moreover, as discussed in the main article, Eq. (S10) with three different bulk-term constants can also be derived directly from 2D symmetry considerations (i.e., isotropy in the *xy* surface), if we assume from the very beginning that the optical field does not vary significantly across *z* within the polymer. Inserting Eq. (S10) into Eq. (S4), we obtain:

$$\Delta h(x,y) = c_1 \partial_k \partial_k \left( E_l^* E_l \right) + c_2 \partial_k \partial_l \left( E_l^* E_k \right) + c_3 \partial_k \partial_k |E_z|^2 + \partial_k \left( c_B E_z^* E_k + c_B^* E_z E_k^* \right) \quad (S11)$$

in which we have introduced the constants

$$c_1 = -\frac{L\tau}{\rho} C_1$$

$$c_2 = -\frac{2L\tau}{\rho} C_2$$

$$c_3 = -\frac{L\tau}{\rho} C_3 \quad (S12)$$

$$c_B = -\frac{\tau}{\rho} C_B$$

In a more explicit way:

$$\Delta h(x,y) = (c_1 + c_2)\left[ \partial_x^2 |E_x|^2 + \partial_y^2 |E_y|^2 \right] + c_1 \left( \partial_x^2 |E_y|^2 + \partial_y^2 |E_x|^2 \right) + c_2 \partial_x \partial_y \left( E_y^* E_x + E_x^* E_y \right)$$
$$+ c_3 \left( \partial_x^2 |E_z|^2 + \partial_y^2 |E_z|^2 \right) + c_B' \left[ \partial_x \mathrm{Re}\left( E_z^* E_x \right) + \partial_y \mathrm{Re}\left( E_z^* E_y \right) \right] + c_B'' \left[ \partial_x \mathrm{Im}\left( E_z^* E_x \right) + \partial_y \mathrm{Im}\left( E_z^* E_y \right) \right] \quad (S13)$$

where

$$c_B' = 2\mathrm{Re}(c_B)$$

$$c_B'' = -2\mathrm{Im}(c_B)$$

The three constants $C_1$, $C_2$, $C_3$ are real by definition, while $C_B$ may be complex in general. If we introduce in the model the additional assumption that all surface currents derive from electric-dipole absorption effects, proportional to $|\boldsymbol{\mu} \cdot \mathbf{E}|^2$, where $\boldsymbol{\mu}$ is a molecular transition-dipole vector, then also $C_B$ must be real and the term with $c_B''$ in Eq. (S13) vanishes, thus leading to Eq. (2) of the main article.

**Simulated material displacement: case of Gaussian-beam illumination.** Figure S2, from S2a to S2g, reports the simulated distributions, at the focal plane of the objective, of each term of equation (2) of the main article, calculated combining the complex values of $E_x$, $E_y$ and $E_z$ obtained at the focal plane for a Gaussian light beam, linearly polarized along the *y*-axis (as for example shown in Figure S1). The terms $\partial_x \mathrm{Re}\left( E_z^* E_x \right)$ and $\partial_y \mathrm{Re}\left( E_z^* E_y \right)$, proportional to the coefficient $c_B$ in equation (2), are omitted because they vanish identically (actually, numerical round-off errors give to these terms random values of the order of $10^{-12}$ relative to the other terms). In fixing the values for the other coefficients of equation (2), the topographical surface modulation resulting from the simulation must be in agreement with that reported many times in the literature[7,8] and also reported in Fig. 2b of the main article, i.e. the two-lobed accumulation along the polarization direction. This profile is compatible with that of the function $\partial_y^2 |E_y|^2$ reported in Figure S2b. In fact, we tested several combinations for the ratios of the coefficients $c_1$, $c_2$ and $c_3$, concluding that both coefficients $c_1$ and $c_3$ should be neglected to best reproduce the experimental profile. Thus, Figures S2h and S2i report respectively the two-and three-dimensional distributions for $\Delta h(x,y)$ as derived by applying equation (2) with coefficients $c_1 = 0$, $c_2 = 1$, $c_3 = 0$. The topographical surface modulation predicted by the simulation is in complete agreement with that reported in literature. It is worth noting that the term in $c_2$, which represents the main component in equation (2) in this illumination conditions, is proportional to the second derivative of the intensity of a Gaussian beam linearly polarized along the *y*-axis with respect to the polarization direction. This, in fact, has motivated the past hypothesis of a mass-transport driving force proportional to the light intensity gradients[6,7,22,31]. A similar analysis can be applied to a circularly polarized Gaussian beam. The two-and three-dimensional distributions for $\Delta h(x,y)$ that we predict in this case are shown in Figures S2l and S2j. Again, these are in excellent agreement with the experimental results reported in the literature[7,8].

**Predicted material displacement: case of two-beam interference illumination.** Furthermore, a configuration often considered in previous works, for large area lithographic structuring of azo-polymers, is that obtained by illumination of the sample surface with the interference pattern of two beams having equal and opposite incidence angles $\theta$ with respect to the normal of the sample surface[5,6,9,10,20]. In particular, the most frequently investigated geometries are with the beam polarizations both parallel to the incidence plane (*xz* plane), called p-p polarization combination, and both perpendicular to the incidence plane, called s-s polarization combination. In the p-p case, the intensity pattern resulting from the interference is proportional to:

$$|E_x|^2 + |E_z|^2 = 2E_0^2 + 2E_0^2 \cos(2\theta)\cos(2kx\sin\theta) \quad (S14)$$

where $E_0$ is the plane-waves' amplitude and $k$ is the wave-number. The resulting polymer relief pattern predicted by equation (2) is instead the following:

$$\begin{aligned}\Delta h &= (c_1+c_2)\partial_x^2|E_x|^2 + c_3\partial_x^2|E_z|^2 = \\ &= -8(c_1+c_2)E_0^2 k^2 \cos^2\theta\sin^2\theta\cos(2kx\sin\theta) + c_3 8 E_0^2 k^2 \sin^4\theta\cos(2kx\sin\theta)\end{aligned} \quad (S15)$$

If the $c_2$ coefficient is dominating, this pattern is π-shifted with respect to the intensity distribution (S14), as indeed observed in the experiments[5,6].

In the s-s polarization combination case the intensity distribution is given by

$$|E_y|^2 = 2E_0^2 + 2E_0^2 \cos(2kx\sin\theta) \quad (S16)$$

and the resulting predicted relief pattern is

$$\Delta h = c_1 \partial_x^2 |E_y|^2 = -8 c_1 E_0^2 k^2 \sin^2\theta\cos(2kx\sin\theta) \quad (S17)$$

This pattern is again π-shifted with respect to the intensity distribution, as observed in the experiments. Moreover, since it is proportional to $c_1$ and we have seen from other cases that this coefficient is much smaller than $c_2$, our model predicts that s-s relief gratings will be much smaller than p-p ones for the same exposure, as also experimentally observed.